\documentstyle[epsfig]{elsart}

\begin{document}

\begin{frontmatter}
\title{Optimal receiver for quantum cryptography\\
with two coherent states}
\author{Konrad Banaszek}
\address{Instytut Fizyki Teoretycznej, Uniwersytet Warszawski,
Ho\.{z}a~69, PL-00-681~Warszawa, Poland}

\begin{abstract}
A setup for discriminating between arbitrary two coherent states of a
single light mode with the highest success rate allowed by quantum
mechanics is presented. Its application to time-multiplexed quantum key
distribution is discussed.\\[\baselineskip]

\end{abstract}
\end{frontmatter}

From the conceptual point of view, the simplest quantum protocol for
establishing a cryptographic key is a transmission of a random sequence
of two non-orthogonal states \cite{B92}. What makes this protocol
secure, is the impossibility of discriminating between these two states
with certainty. Thus, any attempt of eavesdropping inevitably generates
noise in the transmission channel, which can be detected by the
communicating parties \cite{EkerHuttPRA94}. A straightforward optical
realisation of this protocol is the use of two weak coherent states in
an interferometric setup \cite{B92,LANLReview}.

In the two-state protocol, the cryptographic key is constructed
from these events, for which the receiving party managed to identify
unambiguously the obtained quantum state. As the two used states are
non-orthogonal, this procedure can be successful only for a fraction of
the sequence. In order to maximize capacity of the transmission channel,
the recipient should employ a measurement with the smallest possible
probability of yielding an inconclusive result. The general problem of
optimal distinction between two non-orthogonal states has been studied
by Ivanovic \cite{Ivanovic}, Dieks \cite{Dieks}, and Peres \cite{Peres},
who explicitly described a quantum measurement with the required property.
This measurement has been demonstrated for polarisation states of a single
photon \cite{HuttMullPRA96}, and its realisation for a pair of coherent
states with opposite phases has been described \cite{HuttImotPRA95}.

The purpose of this note is to show that the optimal measurement
discriminating between arbitrary two coherent states of a single light
mode can be implemented using a simple optical arrangement. I will
present the quantum optical description of the setup and derive the
positive operator-valued measure representing the measurement in the
infinite-dimensional Hilbert space of the light mode used for
communication. A possible application to time-multiplexed transmission
using a single optical fibre will also be discussed.

\begin{figure}
\centerline{\epsfig{file=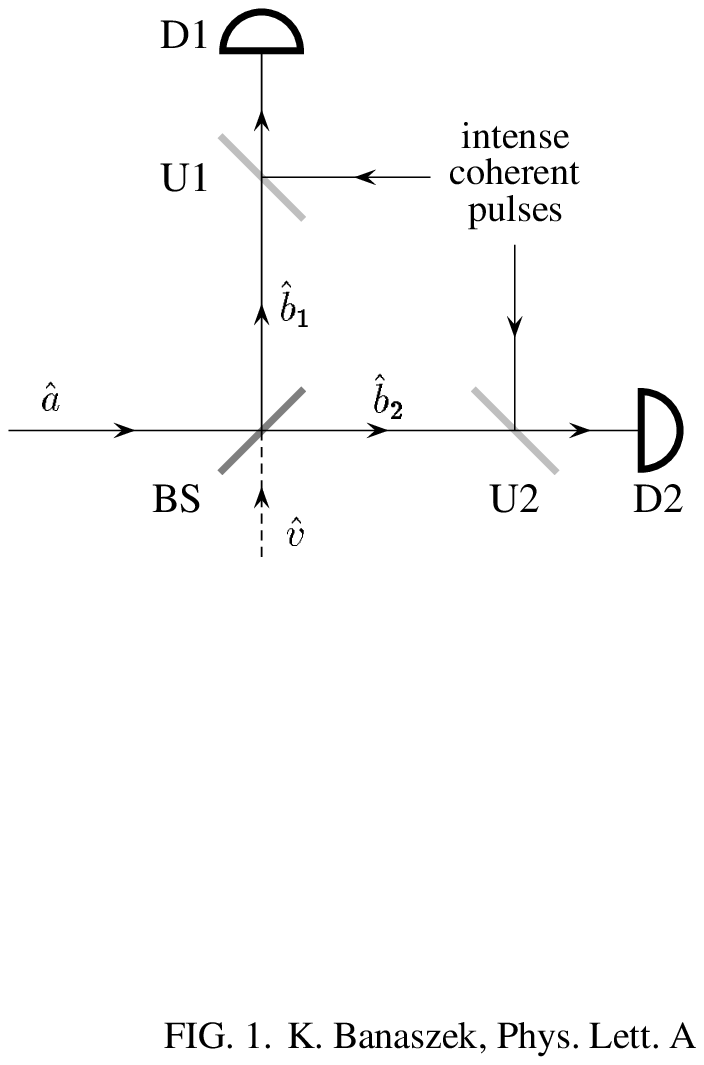,%
bbllx=180,bblly=335,bburx=380,bbury=495,%
clip=}}
\caption{Optical realization of the quantum measurement optimally
discriminating between arbitrary two coherent states of a single light
mode.}
\label{Fig:Setup}
\end{figure}

In each run the sender (traditionally called Alice) generates a pulse
in one of two coherent states: $|\alpha_1\rangle$ or
$|\alpha_2\rangle$. The setup enabling the recipient (known as Bob) to
distinguish optimally between these states is depicted in
Fig.~\ref{Fig:Setup}. First, the received pulse is split using a 50:50
beam splitter BS. The two outgoing fields, described by annihilation
operators $\hat{b}_1$ and $\hat{b}_2$, are linear combinations of the
incident fields:
\begin{equation}
\label{Eq:BS}
\left(
\begin{array}{c} \hat{b}_1 \\ \hat{b}_2 \end{array}
\right)
=
\frac{1}{\sqrt{2}}
\left(
\begin{array}{rr} 1 & 1 \\ 1 & -1 \end{array}
\right)
\left(
\begin{array}{c} \hat{a} \\ \hat{v} \end{array}
\right)
\end{equation}
where $\hat{a}$ is the annihilation operator of the field received from
Alice, and $\hat{v}$ describes the vacuum
field entering through the second
unused port of the beam splitter BS. Each of the outgoing fields
$\hat{b}_1$ and $\hat{b}_2$ is transmitted via an unbalanced beam
splitter (U1 or U2) with a very low reflection coefficient, and an
auxiliary intense coherent pulse injected in the second port. In this
regime, the beam splitters U1 and U2 effectively perform unitary
displacement transformations on the transmitted fields, with the
complex displacement parameters equal to the reflected amplitudes of
the auxiliary coherent fields \cite{BanaWodkPRL96}.

After transmission through the beam splitters U1 and U2, the fields
are measured with the photodetectors D1 and D2, assumed to have
unit quantum efficiency. Each of them
yields one of two results: 0 when no photons were present, and $1$ when
at least one photon was registered. Quantum mechanically, the test
for the presence of photons is described by a pair of projections on
the vacuum state $|0\rangle \langle 0|$ and on the orthogonal subspace
$\hat{I}-|0\rangle\langle 0|$. These projections can be expressed
in terms of the modes $\hat{b}_1$ and $\hat{b}_2$
by applying appropriate displacement operators describing the action of the
beam splitters U1 and U2. This transforms the vacuum states of 
the detector modes into coherent states $|\beta_1\rangle$
and $|\beta_2\rangle $ in the representation
of the modes $\hat{b}_1$ and $\hat{b}_2$.
Thus, the joint measurement performed by Bob is described
by the following two-mode projections describing four possible outcomes:
\begin{eqnarray}
\hat{\cal B}_{00} & = & |\beta_1 \rangle \langle \beta_1|
\otimes |\beta_2 \rangle \langle \beta_2 |
\nonumber \\
\hat{\cal B}_{01} & = & |\beta_1 \rangle \langle \beta_1|
\otimes (\hat{I} - |\beta_2 \rangle \langle \beta_2 | )
\nonumber \\
\hat{\cal B}_{10} & = &
(\hat{I} - |\beta_1 \rangle \langle \beta_1|)
\otimes |\beta_2 \rangle \langle \beta_2 |
\nonumber \\
\hat{\cal B}_{11} & = &
\label{Eq:B}
(\hat{I} - |\beta_1 \rangle \langle \beta_1|)
\otimes (\hat{I} - |\beta_2 \rangle \langle \beta_2 | )
\end{eqnarray}
The parameters of Bob's setup are chosen such that
\begin{equation}
\label{Eq:betai}
\beta_i = \alpha_i/\sqrt{2}, \;\;\;\;\;\;\; i=1,2.
\end{equation}

Let us now find description of Bob's measurement in terms
of the mode $\hat{a}$ used for communication. For
this purpose one has to express the projection operators given
in Eq.~(\ref{Eq:B}) in the representation
of the modes $\hat{a}$ and $\hat{v}$,
and then evaluate the expectation value over the vacuum mode $\hat{v}$.
This yields a positive operator-valued measure
\begin{equation}
\hat{\cal A}_{kl} = \langle 0_v | \hat{\cal B}_{kl} | 0_v \rangle,
\;\;\;\;\;\;\; k,l=0,1
\end{equation}
defined on the Hilbert space of the mode $\hat{a}$.
The calculation described above can be easily performed with
the help of the following
normally ordered representation of projections
on coherent states:
\begin{equation}
|\beta_i \rangle \langle \beta_i |
= \; : \! \exp[-(\hat{b}_i^{\dagger} - \beta_i^\ast)
(\hat{b} - \beta_i)] \! :, \;\;\;\;\;\;\; i=1,2.
\end{equation}
This formula allows one to apply directly the transformation given in
Eq.~(\ref{Eq:BS}) and to eliminate the vacuum mode in a straightforward
manner. The final result is
\begin{eqnarray}
\hat{\cal A}_{00} & = & : \! \hat{Q}_{1} \hat{Q}_{2} \! :
\nonumber \\
\hat{\cal A}_{01} & = & : \! \hat{Q}_{1} \! : - 
: \! \hat{Q}_{1}\hat{Q}_{2} \! :
\nonumber \\
\hat{\cal A}_{10} & = & : \! \hat{Q}_{2} \! : - 
: \! \hat{Q}_{1}\hat{Q}_{2} \! :
\nonumber \\
\hat{\cal A}_{11} & = & \hat{I} - : \! \hat{Q}_{1} \! : - 
: \!\hat{Q}_{2} \! : + : \! \hat{Q}_{1}\hat{Q}_{2} \! :,
\end{eqnarray}
where, after using relation (\ref{Eq:betai})
\begin{eqnarray}
\hat{Q}_{1} & = & \exp\left( - \frac{1}{2}
(\hat{a}^{\dagger} - \alpha^{\ast}_1)(\hat{a} - \alpha_1) \right)
\nonumber \\
\hat{Q}_{2} & = & \exp\left(- \frac{1}{2}
(\hat{a}^{\dagger} - \alpha^{\ast}_2)(\hat{a} - \alpha_2) \right).
\end{eqnarray}

The probabilities of obtaining two results: $00$ and $11$ are
independent of the state sent by Alice:
\begin{equation}
\label{Eq:Inconcl}
\begin{array}{ccccl}
\langle\alpha_1|\hat{\cal A}_{00}|\alpha_1\rangle & = &
\langle\alpha_2| \hat{\cal A}_{00} |\alpha_2\rangle & = & 
\exp(-|\alpha_1 - \alpha_2 |^2/2) \\
\langle\alpha_1 |\hat{\cal A}_{11}|\alpha_1\rangle & = &
\langle\alpha_2| \hat{\cal A}_{11} |\alpha_2\rangle & = & 0,
\end{array}
\end{equation}
whereas the remaining results may occur only for one of the
received states:
\begin{eqnarray}
\langle\alpha_1|\hat{\cal A}_{01}|\alpha_1\rangle & = &
\langle\alpha_2| \hat{\cal A}_{10} |\alpha_2\rangle  = 
1-\exp(-|\alpha_1-\alpha_2 |^2/2) \nonumber \\
\langle\alpha_2| \hat{\cal A}_{01} |\alpha_2\rangle & = & 
\langle\alpha_1|\hat{\cal A}_{10}|\alpha_1\rangle  =  0.
\end{eqnarray}
Thus, the outcomes $01$ and $10$ enable Bob to identify uniquely the
state sent by Alice, and these events can be used to establish a
cryptographic key. From Eq.~(\ref{Eq:Inconcl}) it follows that the
probability of obtaining an inconclusive result is
$\exp(-|\alpha_1-\alpha_2|^2/2)$, and equals the absolute value of the
scalar product $|\langle\alpha_1|\alpha_2\rangle|$. This is the lowest rate
allowed by quantum mechanics \cite{Dieks,Peres}, which acknowledges
that the proposed setup optimally discriminates between the two
coherent states used for secret communication.

The setup presented in Fig.~\ref{Fig:Setup} has a straightforward classical
description. The coherent pulse received from Alice is split into equal
parts and interfered with auxiliary pulses at beam splitters U1 and U2.
These auxiliary pulses are selected such that for the two
possible amplitudes of the pulse generated by Alice, destructive
interference always takes place in one of the arms of the setup. Click
on a detector means that destructive interference 
was not the case in the arm monitored by this detector, which allows
Bob to identify uniquely the amplitude of the received pulse.  
Inconclusive result is when none of the detectors fired. In order to minimize
frequency of such events, Bob should make the non-vanishing field
detected by D1 or D2 as strong as possible. To achieve this, he has
to avoid losses of the field received from Alice,
and use low-reflection beam splitters U1 and U2.

\begin{figure}
\centerline{\epsfig{file=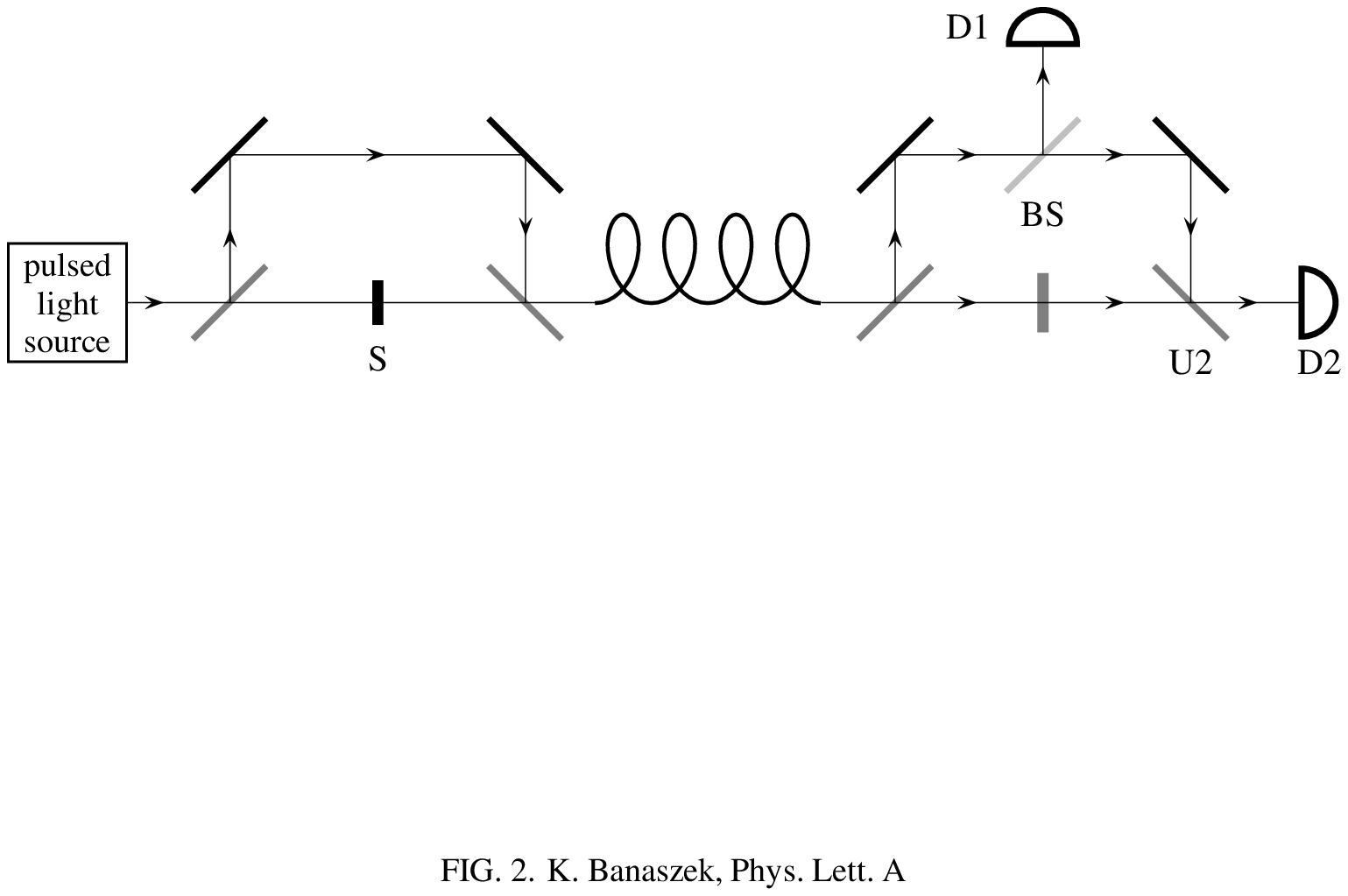,angle=90,clip=,%
bbllx=210,bblly=185,bburx=340,bbury=625,clip=,%
width=\textwidth}}
\caption{A scheme for time-multiplexed quantum key distribution
using the vacuum state and a weak coherent state.}
\label{Fig:Multiplex}
\end{figure}

In practice, the strong coherent pulses necessary to operate the
receiver have to be provided by Alice in order to guarantee fixed phase
difference between fields interfering at unbalanced beam splitters U1
and U2. Furthermore, it is convenient to send the auxiliary pulse along
the same optical path as the signal one with a certain time separation.
This enables stable transmission over large distances using optical
fibers \cite{LANLReview}. A setup for such time-multiplexed quantum key
distribution is shown schematically in Fig.~\ref{Fig:Multiplex}.  Alice's
transmitter is an unequal path interferometer with high-reflecting beam
splitters. On the output of her device, Alice first generates either a
faint coherent pulse or the vacuum by switching the shutter S placed in the
shorter path of the interferometer. This weak quantum signal is followed
by an auxiliary strong pulse travelling via the longer path. Assuming
that all the unbalanced beam splitters in the setup have the same power
transmission $T\ll 1$, the absolute value of the scalar product of the
two states used by Alice is $\exp(-T^2 |\gamma|^2/2)$, where $\gamma$
is the amplitude of the pulse entering Alice's interferometer.

Bob's receiver is an analogous unequal path interferometer. The
unambiguous measurement of Alice's state takes place when the signal
pulse passes the longer path and overlaps with the auxiliary pulse
transmitted via the two high-reflecting beam splitters. In the longer
path, the beam splitter BS extracts about half of the signal to the
detector D1. The remaining part is combined with the auxiliary pulse at
the beam splitter U2. The amplitude and the phase of the auxiliary
field is adjusted in the shorter path so that destructive interference
occurs at the port monitored by the detector D2, when Alice send the
faint coherent pulse.  In this arrangement, the coherent state can
trigger only D1, and the vacuum state can generate a click exclusively
on D2.

In addition to detection events that yield unambiguous information on
the received state, the detectors can also be triggered, with certain
time separation, by the signal transmitted via the shorter path or the
auxiliary pulse travelling via the longer path. The latter possibility is
very likely due to large amplitude of this pulse. Therefore, appropriate
timing between Alice and Bob is necessary in order to define the time
window when the actual measurement takes place.

In order to make detection of two Alice's states equally probable
(assuming that both the detectors are characterized by the same quantum
efficiency $\eta$), the power transmission of the beam splitter BS should
be $1/(2-T)$, which for small $T$ is nearly 50\%. The success rate
then equals $\exp[-\eta(1-T)^2 T^2 |\gamma|^2/(2-T)]$, which for $T\ll
1$ and $\eta \rightarrow 1$ approaches the maximum value allowed by
quantum mechanics.

Let us close this note with a comparison of classical and quantum
descriptions of the presented setup. From the classical point of view,
the reason for inconclusive results is that measured fields are too
weak to trigger always one of the photodetectors. It may therefore seem
that what primarily determines the failure rate is the characteristics
of the photodetectors. However, quantum theory of radiation reveals
that the minimum failure rate is an intrinsic property of the two
states of light to be distinguished, and depends solely on their scalar
product. Another difference between classical and quantum descriptions
is the role of vacuum fields. Classically, the vacuum field entering
the input port of the beam splitter BS is irrelevant; in the quantum
mechanical picture, it plays the role of an ``ancilla'' system, which
extends the positive operator-valued measure $\hat{\cal A}_{kl}$ to the
projective measure $\hat{\cal B}_{kl}$ on the two-mode Hilbert space.

\section*{Acknowledgements}

The author wishes to acknowledge discussions with Prof.\ K.
W\'{o}dkiewicz. This research was supported by the Polish KBN grant
No.\ 2P03B~013~15 and by Stypendium Krajowe dla M{\l}odych
Naukowc\'{o}w Fundacji na rzecz Nauki Polskiej.


\begin{thebibliography}{9}

\bibitem{B92}
C. H. Bennett, Phys.\ Rev.\ Lett.\ 68 (1992) 3121.

\bibitem{EkerHuttPRA94}
A. K. Ekert, B. Huttner, G. M. Palma, and A. Peres, Phys.\ Rev.\ A
50 (1994) 1047.

\bibitem{LANLReview}
R. J. Hughes {\em et al.\/}, Contemp.\ Phys.\ 36 (1995) 149;
S. J. D. Phoenix and P. D. Townsend, Contemp.\ Phys.\ 36 (1995) 165.

\bibitem{Ivanovic}
I. D. Ivanovic, Phys.\ Lett.\ A 123 (1987) 257.

\bibitem{Dieks}
D. Dieks, Phys. Lett.\ A 126 (1988) 303.

\bibitem{Peres}
A. Peres, Phys.\ Lett.\ A 128 (1988) 19.

\bibitem{HuttMullPRA96}
B. Huttner, A. Muller, J. D. Gautier, H. Zbinden, and N. Gisin,
Phys.\ Rev.\ A 54 (1996) 3783.

\bibitem{HuttImotPRA95}
B. Huttner, N. Imoto, N. Gisin, and T. Mor, Phys.\ Rev.\ A
51 (1995) 1863.

\bibitem{BanaWodkPRL96}
K. Banaszek and K. W\'{o}dkiewicz, Phys.\ Rev.\ Lett.\ 76 (1996) 4344.

\end{thebibliography}
\end{document}